\documentclass[aps,prd,preprint,superscriptaddress,tightenlines,nofootinbib]{revtex4} 
\usepackage{epsfig}
\usepackage{graphicx}
\usepackage{dcolumn}
\usepackage{bm}

\newcommand{\jpsi}{J/ \psi}

\def\upsl1{ \Upsilon {\rm (1S)}}

\def\ups4s{ \Upsilon (4S)}

\def\ipb{{\rm pb^{-1}}}

\def\mbc{M_{\rm bc}}

\def\de{\Delta E}

\def\y5s{\Upsilon (5S)}
\def\bstbst{B^*\bar{B}^*}
\def\bsstbsst{B_s^{(*)}\bar{B}_s^{(*)}}
\def\bbbar{B\bar{B}}
\def\bbstar{B\bar{B}^{*}}
\def\bbpipi{B\bar{B}\pi\pi}
\def\bother{B^{(*)}\bar{B}^{(*)}\pi}
\def\dkpi{D^0\to K^-\pi^+}
\def\dkpipi0{D^0\to K^-\pi^+\pi^0}
\def\dk3pi{D^0\to K^-\pi^+\pi^+\pi^-}
\def\dktwopi{D^+\to K^-\pi^+\pi^+}
\def\kpi{K^-\pi^+}
\def\kpipi0{K^-\pi^+\pi^0}
\def\k3pi{K^-\pi^+\pi^+\pi^-}
\def\ktwopi{K^-\pi^+\pi^+}
\def\dstpd0pi{D^{*+}\to D^0\pi^+}
\def\dst0d0pi0{D^{*0}\to D^0\pi^0}
\def\dstpdppi0{D^{*+}\to D^+\pi^0}

\begin{document}

\preprint{\tighten\vbox{\hbox{\hfil CLNS 06-1950}
                        \hbox{\hfil CLEO 06-1}
}}

\title{
First Measurements of the Exclusive Decays of the $\y5s$ to B Meson Final States
and Improved $B_s^*$ Mass Measurement}

\author{O.~Aquines}
\author{Z.~Li}
\author{A.~Lopez}
\author{H.~Mendez}
\author{J.~Ramirez}
\affiliation{University of Puerto Rico, Mayaguez, Puerto Rico 00681}
\author{G.~S.~Huang}
\author{D.~H.~Miller}
\author{V.~Pavlunin}
\author{B.~Sanghi}
\author{I.~P.~J.~Shipsey}
\author{B.~Xin}
\affiliation{Purdue University, West Lafayette, Indiana 47907}
% ...
\collaboration{CLEO Collaboration} 
\noaffiliation

\date{Mar. 17, 2006}

\begin{abstract}
	Using 420 $\ipb$ of data collected on the $\y5s$ resonance 
with the CLEO III detector, we
reconstruct $B$ mesons in 25 exclusive decay channels to
measure or set upper limits on the decay rate of $\y5s$ into 
B meson final states. 
We measure the inclusive $B$ cross-section to
be $\sigma(\y5s\to\bbbar(X))=(0.177\pm 0.030\pm 0.016)$ nb 
and make the first measurements of the production rates 
of $\sigma(\y5s\to\bstbst)=(0.131\pm 0.025\pm 0.014)$ nb and
$\sigma(\y5s\to\bbstar)=(0.043\pm 0.016\pm 0.006)$ nb, respectively.
We set 90\% confidence level limits of
$\sigma(\y5s\to B\bar{B})<0.038$ nb,
$\sigma(\y5s\to\bother)<0.055$ nb and
$\sigma(\y5s\to\bbpipi)<0.024$ nb.
We also extract the most precise value of the $B_s^*$ mass to date, 
$M(B_s^*)=(5411.7\pm1.6\pm0.6)$~MeV/$c^2$.
\end{abstract}

\maketitle

\setcounter{footnote}{0}

\newpage

	The $\y5s$ resonance was discovered by the CLEO~\cite{cleo_y5s} and 
CUSB~\cite{cusb_y5s} collaborations. Its production cross-section and mass were 
measured to be about 0.35~nb and (10.865$\pm$0.008)~GeV/$c^2$~\cite{cleo_y5s}, 
respectively. Final states can be: $B\bar{B}$, $\bbstar$, $\bstbst$, 
$B\bar{B}\pi$, $\bbstar\pi$, $\bstbst\pi$, $\bbpipi$,
$B_s\bar{B_s}$, $B_s\bar{B_s^*}$ and $B_s^*\bar{B}_s^*$. Here, 
$B$=$B_u$ or $B_d$, and the $\pi$ may be charged or neutral 
(consistent with charge zero of the final state). 
Throughout this article, we use $\bbstar$ to signify both $\bbstar$ and $B^*\bar{B}$.
Including a symbol in parentheses indicates that  it may or may not be present.
The $B$ cross-section in this region is well-described by the
Unitarized Quark Model (UQM)~\cite{uqm}, which 
predicts that about 1/3 of the $b\bar{b}$ decay rate is
to $B_s^{(*)}\bar{B}_s^{(*)}$ and that $\bstbst$ dominates the inclusive $B$ rate.
A previous CLEO measurement using inclusive $D_s$ production 
revealed that $\y5s\to\bsstbsst$ constitutes (16.0$\pm$2.6$\pm$5.8)\% 
of the total $b\bar{b}$ rate~\cite{radia_5s}. A second analysis~\cite{victor_5s},
which performed exclusive reconstruction of $B_s$ mesons found
$\sigma(e^+e^-\to B_s^*\bar{B}_s^*)=(0.11^{+0.04}_{-0.03}\pm0.02)$ nb
(about 1/3 of the total hadronic resonance cross-section). 
The two results are consistent with each other and with predictions 
of the UQM.

	In this Letter, we measure the contributions of various
$B$ meson final states to the $\y5s$ decay. These measurements may better
constrain coupled-channel models in the $\Upsilon$ mass region as
well as near the lower $\psi$ resonances~\cite{eichten}.
We also exploit exclusively reconstructed $B$ mesons from this analysis
and the corresponding $B_s$ analysis~\cite{victor_5s} to extract
the most precise measurement of the $B^*_s$ mass to date.


	CLEO III is a general purpose solenoidal detector
that includes a tracking system for measuring momenta and specific ionization ($dE/dx$)
of charged particles, a Ring Imaging Cherenkov detector (RICH) to aid in particle
identification, a CsI calorimeter for detection of electromagnetic showers, and
a muon system for identifying muons~\cite{cleo3}.

	The analysis presented here uses 420 $\ipb$ of data collected on the
$\y5s$ resonance ($\sqrt{s}=10.868$~GeV) at the Cornell Electron Storage Ring.
Using techniques pioneered at the $\ups4s$, we utilize two kinematic
variables: the energy difference $\de\equiv~E_{\rm beam}-E_{B}$ and the
beam-constrained mass $\mbc\equiv\sqrt{E_{\rm beam}^2-\vec{p}_{B}^2}$, 
where $E_{B}$ ($\vec{p}_{B}$) is the energy (momentum) of the reconstructed 
$B$ candidate and $E_{\rm beam}$ is the beam energy. Because of its low energy,
reconstruction of the photon in
$B^*\to~B\gamma$ is not essential; to maintain high efficiency,
we do not reconstruct the $B^*$. (Charge conjugate final states are implied 
throughout this Letter.)

	To obtain a $B$~meson sample of high purity, events are required
to have at least five charged tracks and a ratio of the second to zeroth
Fox-Wolfram moment~\cite{fox-wolfram}, $R_2<0.25$. 
Candidate $B$~mesons are reconstructed in exclusive final states
containing either $\jpsi$ or $D^{(*)}$ mesons.

	Charged particles are required to pass standard
selection criteria and are identified by using
their measured momenta in conjunction with $dE/dx$, RICH, calorimeter
and muon system information. For particle types $i,~j$ ($i,~j=\pi,K,p)$ 
we define $\chi^2$-like quantities for $dE/dx$ as the difference
in the measured and expected $dE/dx$, normalized by the uncertainty,
{\it i.e.,} $\chi_i^{dE/dx}\equiv (dE/dx^{\mathrm{meas}}_i-dE/dx^{\mathrm{exp}}_i)/\sigma_i$,
and for RICH as
$\Delta\chi^2_{i,j}\equiv {\cal{L}}_{i}-{\cal{L}}_{j}$ (difference in negative log-likelihood 
between hypotheses $i$ and $j$), respectively. We require
at least 3 detected Cherenkov photons from the RICH.
Pions are identified by requiring $|\chi_{\pi}^{dE/dx}|<4$ or $\Delta\chi^2_{\pi,K}<5$.
For kaons, we define a combined quantity,
$\chi^2_{\mathrm{comb}}\equiv (\chi_K^{dE/dx})^2-(\chi_{\pi}^{dE/dx})^2+{\Delta\chi^2_{K,\pi}}$ and
require $\chi^2_{\rm comb}<0$. Electron candidates are formed from
particles that have a ratio of calorimeter energy ($E_e$) to measured
momentum ($p_e$) in the range $0.5<E_e/p_e<1.25$. Muons are identified by 
either having penetrated at
least 3 layers of iron absorber or by having deposited energy in the
calorimeter consistent with a minimum ionizing particle ($E<300$~MeV).
Photons are formed from showers that have deposited at least 30~MeV of energy in
the calorimeter and are not associated with a charged track. Pairs of photons 
that have an invariant mass within 2 standard deviations ($\sigma\sim6$~MeV/$c^2$) of the 
known $\pi^0$ mass ($M_{\pi^0}$)~\cite{pdg} are defined as $\pi^0$ candidates and 
are kinematically constrained to give $M_{\pi^0}$.

	Candidate $\jpsi$'s are formed from $\mu^+\mu^-$ or $e^+e^-$ pairs.
For muon pairs, we require $3.05<M_{\mu^+\mu^-}<3.14$~GeV/$c^2$.
For $e^+e^-$ combinations with $1.50<M_{e^+e^-}<3.14$~GeV/$c^2$,  
bremsstrahlung photons are searched for among the showers with no
matching charged track and within a $5^{\circ}$ cone about each electron's 
initial direction. For each $\mu^+\mu^-$ and $e^+(\gamma)e^-(\gamma)$ candidate, we 
perform a mass-constrained fit to the $\jpsi$ mass~\cite{pdg} and
make a loose requirement that the fit $\chi^2$ per degree of freedom is less than 100. 
Candidate $\rho^+$ ($K^0_S$) [$K^{*0}$] mesons are formed from
$\pi^+\pi^0$ ($\pi^+\pi^-$) [$K^+\pi^-$] combinations that have
an invariant mass in the range from 620-920 (490-505)
[820-970]~MeV/$c^2$. $D^+$ ($D^0$) meson candidates are
reconstructed via their decays to $\ktwopi$ ($\kpi$, $\k3pi$ and 
$\kpipi0$) and are required to have an invariant
mass within 2$\sigma$ of their PDG~\cite{pdg} values.
To reduce combinatorial background in $\dkpipi0$, we require $p_{\pi^0}>400$~MeV/$c$. 
Candidate $D^{*+}\to~D^0\pi^+$ ($D^{*+}\to~D^+\pi^0$) decays are formed from 
$D$ and $\pi$ candidates that have a mass difference in the range 
$144<M_{D^{*+}}-M_{D^0}<147$~MeV/$c^2$ ($139<M_{D^{*+}}-M_{D^+}<143$~MeV/$c^2$). 
Similarly, $D^{*0}$ mesons are reconstructed in $D^0\pi^0$, and the mass 
difference is required to be in the interval $140<M_{D^{*0}}-M_{D^0}<144$~MeV/$c^2$.

	Candidate $B$ mesons are reconstructed in the 25 decay channels listed 
in Table~\ref{tab:modes}. For $B\to~D\rho$ and $B\to~D^*\rho$~\cite{b2dstrho}, 
we take advantage of the helicity angle ($\theta_h$)~\cite{hel_angle} distribution in 
these decays and require $|\cos\theta_h|>0.3$.
To improve the signal-to-background ratio, we also reject low momentum (backward-emitted)
$\pi^0$'s from the $\rho^+$ decays by requiring $\cos\theta_h>-0.7$. 
Table~\ref{tab:modes} also shows the product branching fractions, ${\cal{B}}_i$,
including the branching ratios of the daughter modes~\cite{pdg}, and the reconstruction
efficiencies, $\epsilon_i$ determined from Monte Carlo simulations~\cite{pythia,qq,photos}
of these decays followed by a {\sc geant}~\cite{geant} 
based detector simulation. 
We validate our simulation
and analysis procedure by measuring branching fractions for these decay modes
using data collected on the $\ups4s$ resonance. Good agreement
with the world averages are found for all modes.

\begin{table*}[hbt] 
\begin{center}  
\caption{Modes used in exclusive $B$ meson reconstruction along with
their product branching fractions (${\cal{B}}_i$)~\cite{pdg}, reconstruction 
efficiencies ($\epsilon_i$),
and expected yields ($N_{\rm exp}^i$) in data assuming  $\sigma(\y5s\to\bbbar(X))=0.2$ nb.
\label{tab:modes}}
\begin{tabular}{lccc}\hline\hline 
Mode $i$ & ${\cal{B}}_i$ & $\epsilon_i$ & $N_{\rm exp}^i$ \\
     & $(10^{-4})$ &   (\%)     &  	    \\
\hline
$B^+\to\jpsi K^+$      			   & $~~1.18\pm0.05~~$  &  43.4$\pm$0.9  & 4.3  \\
$B^0\to\jpsi K^{*0}$, $K^{*0}\to K^+\pi^-$ & $1.03\pm0.08$  &  25.6$\pm$0.6  & 2.2  \\
$B^0\to\jpsi K^0_S$    			   & $0.34\pm0.02$  &  37.2$\pm$1.7  & 1.1 \\
\hline\hline
$B^-\to D^0\pi^-$, $\dkpi$ & $1.87\pm0.09$    & 34.4$\pm$0.5	& 5.4   \\
$B^-\to D^0\pi^-$, $\dkpipi0$ & $6.48\pm0.56$ & 12.8$\pm$0.5    & 7.0   \\
$B^-\to D^0\pi^-$, $\dk3pi$ & $3.67\pm0.22$   & 20.7$\pm$0.7    & 6.4   \\
\hline
$B^-\to D^{*0}\pi^-$, $\dst0d0pi0$, $\dkpi$ & $1.08\pm0.11$    & 11.1$\pm$0.5 	& 1.0  \\
$B^-\to D^{*0}\pi^-$, $\dst0d0pi0$, $\dkpipi0$ & $3.76\pm0.47$ & 2.5$\pm$0.2	& 0.8  \\
$B^-\to D^{*0}\pi^-$, $\dst0d0pi0$, $\dk3pi$ & $2.13\pm0.23$   & 6.9$\pm$0.4	& 1.2  \\
\hline
$B^-\to D^0\rho^-$, $\dkpi$       & $5.11\pm0.14$ &    8.2$\pm$0.3 & 3.5  \\
$B^-\to D^0\rho^-$, $\dkpipi0$    & $17.69\pm1.36$ &   3.0$\pm$0.2 & 4.5  \\
$B^-\to D^0\rho^-$, $\dk3pi$      & $10.02\pm0.42$ &   5.2$\pm$0.3 & 4.4  \\
\hline
$B^-\to D^{*0}\rho^-$, $\dst0d0pi0$, $\dkpi$ & $2.31\pm0.42$    & 2.1$\pm$0.1 & 0.4 \\
$B^-\to D^{*0}\rho^-$, $\dst0d0pi0$, $\dkpipi0$ & $7.90\pm1.54$ & 0.7$\pm$0.1 & 0.5\\
$B^-\to D^{*0}\rho^-$, $\dst0d0pi0$, $\dk3pi$ & $4.56\pm0.84$   & 1.5$\pm$0.1 & 0.6\\
\hline
$B^0\to D^+\pi^-$, $\dktwopi$      & $2.64\pm0.25$ &   30.9$\pm$1.5 &  6.9 \\
\hline
$B^0\to D^{*+}\pi^-$, $\dstpd0pi$, $\dkpi$ & $0.71\pm0.06$    & 22.0$\pm$0.4 & 1.3   \\
$B^0\to D^{*+}\pi^-$, $\dstpd0pi$, $\dkpipi0$ & $2.47\pm0.27$ & 4.0$\pm$0.1  & 0.8   \\
$B^0\to D^{*+}\pi^-$, $\dstpd0pi$, $\dk3pi~~$ & $1.40\pm0.12$   & $~~12.2\pm0.4~~$ & 1.4  \\
\hline
$B^0\to D^{*+}\pi^-$,$\dstpdppi0$,$\dktwopi$ & $0.78\pm0.08$ & 7.3$\pm$0.5   & 0.5  \\
\hline
$B^0\to D^+\rho^-$, $\dktwopi$ & $7.08\pm1.28$ &  6.6$\pm$0.4  &    3.9  \\
\hline
$B^0\to D^{*+}\rho^-$, $\dstpd0pi$, $\dkpi$& $1.75\pm0.24$    & 4.3$\pm$0.2 & 0.6  \\
$B^0\to D^{*+}\rho^-$, $\dstpd0pi$, $\dkpipi0$& $6.08\pm0.93$ & 1.3$\pm$0.1 & 0.7  \\
$B^0\to D^{*+}\rho^-$, $\dstpd0pi$, $\dk3pi$& $3.44\pm0.48$   & 2.7$\pm$0.2 & 0.8  \\
\hline
$B^0\to D^{*+}\rho^-$,$\dstpdppi0$, $\dktwopi$ & $1.94\pm0.29$ & 1.7$\pm$0.1 & 0.3  \\
\hline\hline
Total &\multicolumn{2}{c}{$\sum{{\cal{B}}_i\epsilon_i}=7.2\times10^{-4}$} & 60.4  \\  
\hline\hline
\end{tabular}   
\end{center} 
\end{table*}

\begin{figure}
\centerline{
\includegraphics[width=3.25in]{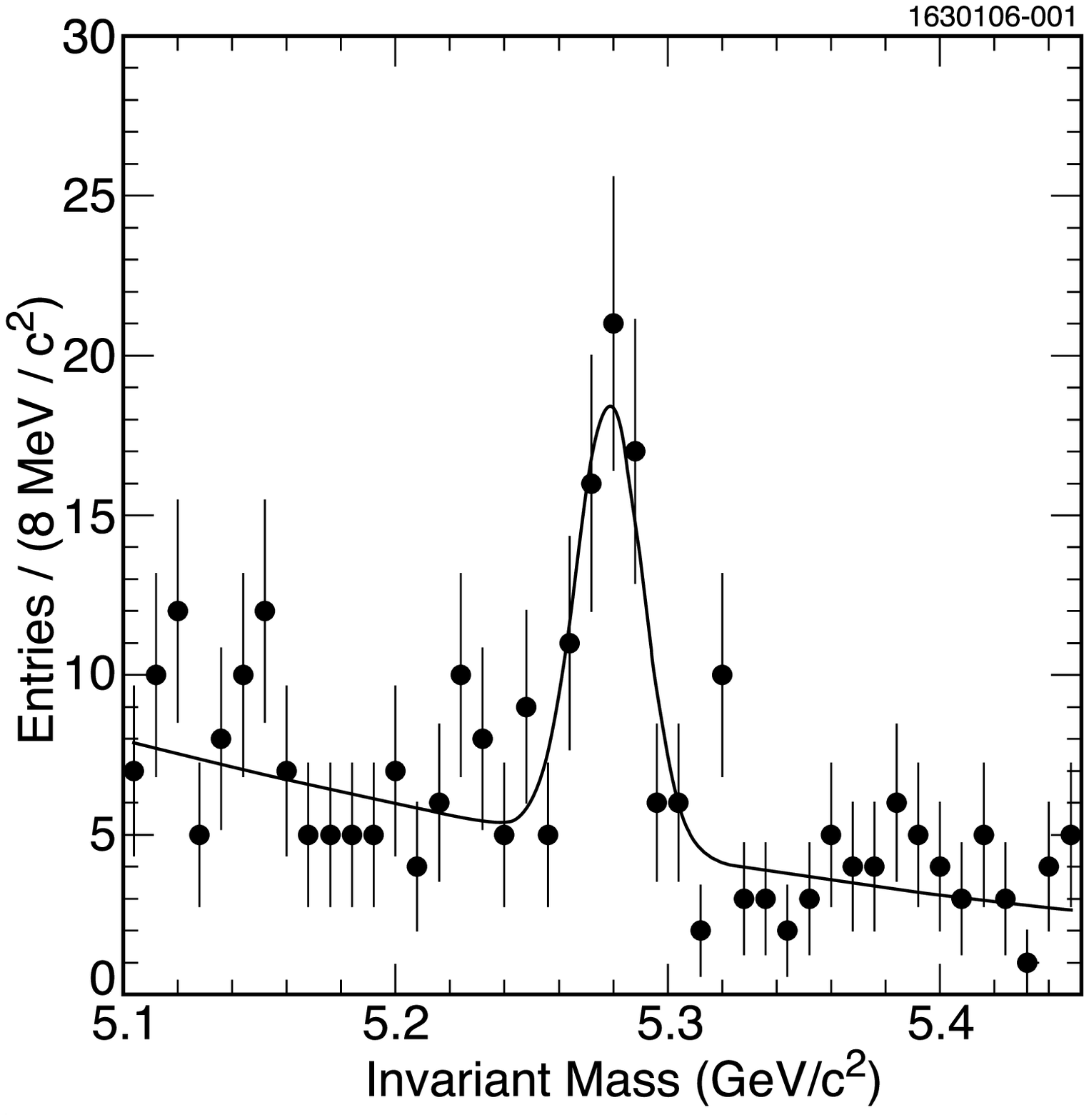}}
\caption{\label{fig:b2all_invmass} 
The $B$ meson invariant mass distribution for all $B$ decay modes
listed in Table~\ref{tab:modes} in $\y5s$ data.
The points are the data and the curve is the fit described in the text.}
\end{figure}

	We first determine the total $B$ meson yield by fitting the invariant mass 
distribution formed from candidates in the $\mbc$, $\de$ region of $5.272<\mbc<5.448$~GeV/$c^2$, 
$-0.2<\de < 0.45$~GeV. The relatively wide $\de$ region is used to avoid biasing
the background shape.
The invariant mass distribution, shown in Fig.~\ref{fig:b2all_invmass},
is fit to the sum of a second-order polynomial background and a Gaussian signal shape 
whose width is 
fixed to 12.3~MeV/$c^2$, the expected average resolution of these candidates.
We find a yield of $53.2\pm9.0$ events; fitting to the $\jpsi$ and $D^{(*)}$ 
distributions individually results in 11.2$\pm$3.5 
and 42.3$\pm$8.4 $D^{(*)}$ events, respectively.
Using $\sum{{\cal{B}}_i\epsilon_i}=7.2\times10^{-4}$ (see Table~\ref{tab:modes}),
we measure a cross-section $\sigma(\y5s\to\bbbar(X))=(0.177\pm0.030)$~nb.

\begin{figure}
\centerline{
\includegraphics[width=3.25in]{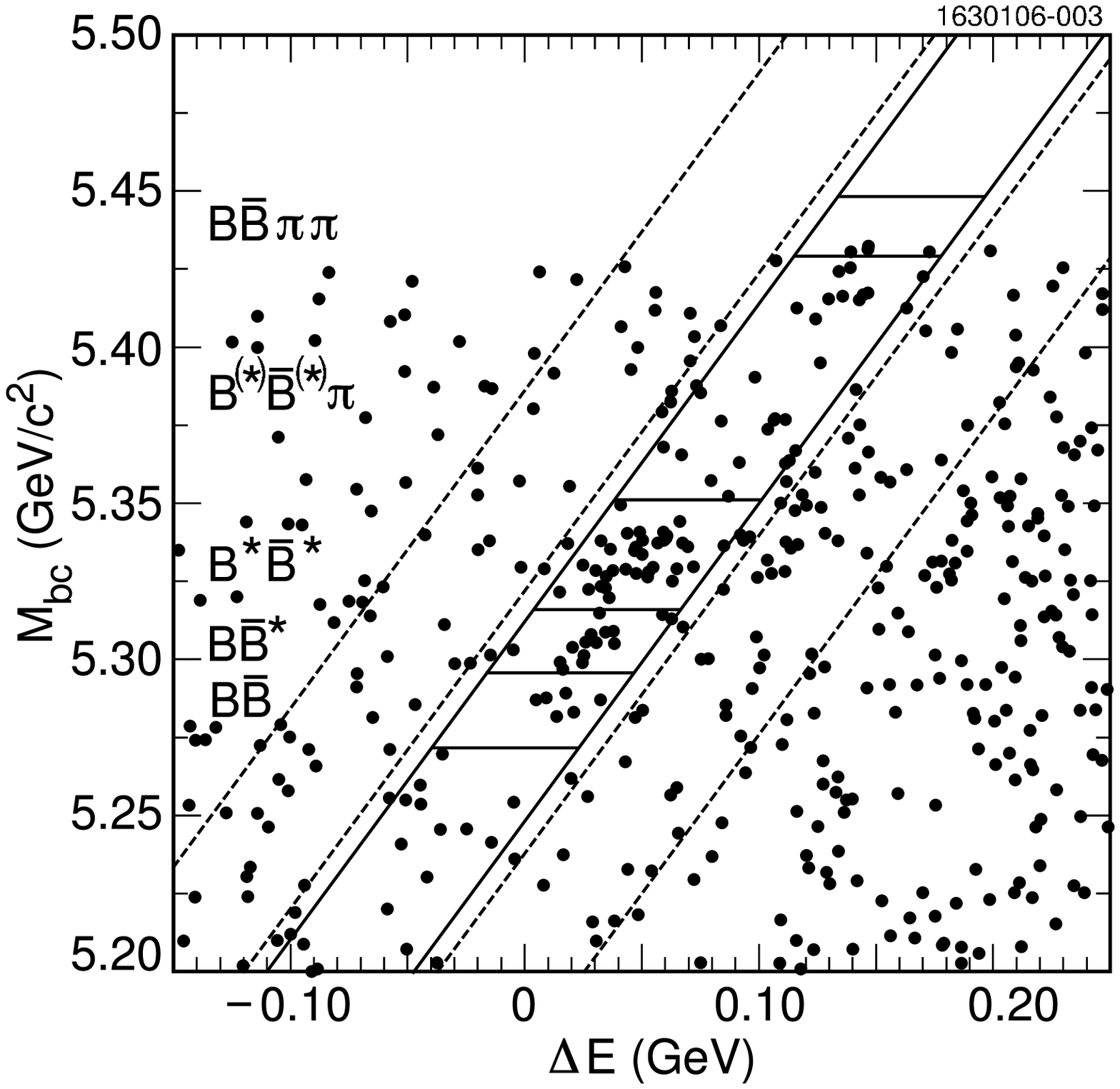}}
\caption{\label{fig:b2all_mbc_vs_de} 
Scatter plots of $\mbc$ {\it vs.} $\de$ for all $B$ decay modes
listed in Table~\ref{tab:modes} in $\y5s$ data. The diagonal lines show the expected signal
(solid) and 2 sideband (dashed) regions, as discussed in the text. The horizontal
lines show the regions for the various $\bbbar(X)$ final states.}
\end{figure}

	Figure~\ref{fig:b2all_mbc_vs_de} shows the reconstructed events in the
$\mbc-\de$ plane for $\y5s$ data. Signal and sideband regions are defined
using MC simulations of these final states. To extract rates for $\bbbar$,
$\bbstar$, and $\bstbst$ separately, we select events in a signal region defined by the
area between the diagonals $\mbc=1.018\de+5.248$~GeV/$c^2$ and $\mbc=1.018\de+5.312$~GeV/$c^2$.
This restricted signal
region has a total $\sum{{\cal{B}}_i\epsilon_i}=5.7\times10^{-4}$.
Lower and upper sidebands of the same $\de$ width, also shown in 
Fig.~\ref{fig:b2all_mbc_vs_de}, are shifted to the left and right
of the signal region by 10~MeV, respectively.  

	The $\bbbar$, $\bbstar$ and $\bstbst$ final states are kinematically 
well separated, but $\bother$ final states have a large degree of overlap, 
and with the limited statistics cannot be distinguished. The
$\bbpipi$ final states, because of the limited phase space, 
peak at $\mbc\simeq~E_{\rm beam}$. If their rate is large enough, their
shape (in $\mbc$) will be sufficient to distinguish them from the broad
tail of $\bother$ final states that extend into the $\mbc$ region of $\bbpipi$.

\begin{figure}
\centerline{
\includegraphics[width=3.25in]{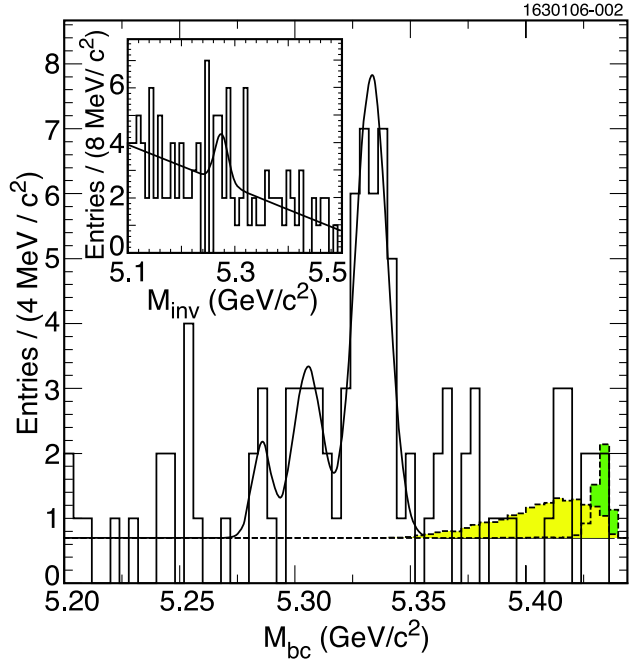}}
\caption{\label{fig:b2all_mbc} 
Distribution of $\mbc$ for all reconstructed $B$ modes
listed in Table~\ref{tab:modes} for $\y5s$ data.
The histogram displays the data, the curve shows the fit described in the text
and the flat line shows the background as determined from a fit to the sidebands
as discussed in the text.
Distributions for $\bother$ (lightly shaded) and $\bbpipi$ (darker shading) obtained
from MC simulation of these final states are superimposed for illustrative purposes.
The inset shows the invariant mass of candidates in the $\bother$ $\mbc$ region
with the fit superimposed.}
\end{figure}

	Events in the signal
region of Fig.~\ref{fig:b2all_mbc_vs_de} are projected onto the $\mbc$ axis
(see Fig.~\ref{fig:b2all_mbc}) and fit to the sum of a flat background and 
three Gaussian signals, one each for the $\mbc$ peaks produced by
$\bbbar$, $\bbstar$ and $\bstbst$ events. 
The signal resolutions are set to $\sigma=$4.0, 6.2 and 7.0~MeV/$c^2$, 
respectively, as determined from MC simulation, and the background is 
fixed to 0.7 events/4~MeV, as determined from the average of the 
upper and lower sidebands. In the fit we use the precisely known
mass difference $M_{B^*}-M_B$~\cite{pdg} and constrain its value
to 47.5~MeV/$c^2$~\cite{mdif}. The middle peak, which corresponds to
$\y5s\to\bbstar$, is not constrained in the fit
and is found to be within 1$\sigma$ of the expected value.

	The fitted yields are $3.7^{+3.1}_{-2.4}$ $\bbbar$, $10.3\pm3.9$ $\bbstar$
and 31.4$\pm$6.1 $\bstbst$ events. Only the latter two are statistically significant
with significances, determined from the change in log-likelihood when
the contribution from each peak is removed, of 4.3$\sigma$ and 7.6$\sigma$, respectively. 
For $\bbbar$, we compute an upper limit of 7.5 events at 90\% confidence level (CL).
A potential excess in $\bother$ is examined
by plotting the invariant mass of candidates in the $\bother$ region defined
by $5.351<\mbc<5.429$~GeV/$c^2$ and $-0.2<\de < 0.45$~GeV, which should
exhibit a peak at $M_B$ (see inset in Fig.~\ref{fig:b2all_mbc}). 
This $\mbc-\de$ region includes ($88\pm6$)\% of reconstructed $\bother$ events,
where the uncertainty reflects the maximum variation based on the
possible $\bother$ final states. That distribution is fit to the sum of a 
Gaussian signal whose mean and r.m.s width are constrained to 5.279~GeV/$c^2$
and 12.3~MeV/$c^2$, respectively, and a linear background shape.
The yield of $6.7^{+5.1}_{-4.5}$ is not statistically significant, and we 
compute an upper limit of 13.1 events at 90\% CL.

	For the $\bbpipi$ final state, we select events in the region
$5.429<\mbc<E_{\rm beam}$~GeV/$c^2$ and $-0.2<\de < 0.45$~GeV and find
three events consistent with $M_B$. This additional requirement on $\mbc$ and $\de$
has an efficiency of $(95\pm2)\%$. While the combinatorial background is
small ($\sim0.3$ events), the cross-feed from $\bother$ into the $\bbpipi$ signal
region is ($12\pm6$)\%. If we conservatively assume a $\bother$ yield equal to
its 90\% upper limit value and that the 3 $\bbpipi$ candidates are also
$\bother$, we would expect 1.0 - 2.9 $\bother$ events to lie within the
$\bbpipi$ mass region. Based on this range of expected background and
3 observed events, we take the most conservative upper limit 
on $\y5s\to\bbpipi$, which corresponds to $6.4$~\cite{feldman} events at 90\% CL.
For illustration, we superimpose on Fig.~\ref{fig:b2all_mbc}, 6.7 $\bother$ (lightly
shaded, with a ratio of 1:1:1 ratio for $\bstbst\pi$:$\bbstar\pi$:$\bbbar\pi$) and 
3 $\bbpipi$ (darker shading) events. 
Yields, efficiencies, cross-sections and relative production fractions are 
summarized in Table~\ref{tab:summary}. We also show the cross-sections as
determined from the $\jpsi$ and $D^{(*)}$ modes separately.
We find that $\bstbst$ is indeed dominant, comprising ($74\pm15)$\% of the $\bbbar(X)$ rate. 

	Several sources of systematic uncertainty on the cross-sections
measurements are considered.
Potential errors from the background normalization and shape are evaluated by 
using different background parameterizations and varying the normalization within 
its uncertainty. The corresponding uncertainties in the cross-sections vary from
3.1\% for $\y5s\to\bbbar(X)$ to 16.7\% for $\bbbar$. Uncertainties in the reconstruction
efficiencies include contributions from charged particle tracking and identification, 
$K^0_s$ and $\pi^0$ reconstruction, and finite MC statistics. Averaged over all modes, 
we find an uncertainty of 6.5\%.
The analysis procedure was also checked by comparing $B$-meson branching fractions 
in our signal modes measured using data collected on the $\ups4s$ resonance with
PDG values~\cite{pdg}. We find a relative difference of ($1\pm3$)\%, 
averaged over all modes, indicating that the efficiencies are well understood.
Errors due to the fixed signal shape parameters are determined
by varying them within their uncertainties and refitting (3\%-4\%). The occurrence
of multiple candidates in data (in a single event) are found to agree with simulation 
to within 3\%. Uncertainties on input branching fractions and measured 
integrated luminosity contribute 3\% and
2\%, respectively. These systematic uncertainties are added in quadrature
and the resulting values are included in the cross-sections shown in 
Table~\ref{tab:summary}.

\begin{table*}[hbt] 
\begin{center}  
\caption{Summary of yields, efficiencies, cross-sections and fractional contributions
of various subprocesses in $\sigma(\y5s\to\bbbar(X))$ decays.
Upper limits are set at the 90\% CL.
Uncertainties are from statistical and systematic sources, respectively.
\label{tab:summary}}
\begin{tabular}{lcccc}\hline\hline
$\y5s$    	&        Yield    & Efficiency & Cross-Section & $\sigma/\sigma(\y5s\to\bbbar(X))$  \\
$~~\to$		& (\#Events)	  &$(10^{-4})$ &       (nb)    &     (\%)      \\
\hline
$B\bar{B}$  		&  $<7.5$       & $5.7\pm0.4$ & $<0.038$ 	  	  & 22 \\
$B^*\bar{B}$  		&  ~$10.3\pm3.9$~ & ~$5.7\pm0.4$~ & ~$0.043\pm0.016\pm0.006$~ & $24\pm9\pm3$ \\
$B^*\bar{B}^*$  	&  $31.4\pm6.1$ & $5.7\pm0.4$ & $0.131\pm0.025\pm0.014$ & $74\pm15\pm8$ \\
$\bother$	  	&  $<13.1$ 	& $6.3\pm0.6$ & $<0.055$ 	  & $<32$ \\
$\bbpipi$		&  $<6.4$ 	& $6.8\pm0.5$ & $<0.024$ 	  & $<14$ \\
\hline
$\sigma(\y5s\to\bbbar(X))$ &  53.2$\pm$9.1   & $7.2\pm0.5$ & $0.177\pm0.030\pm0.016$ & \\
\hline\hline
$\jpsi$ Modes		&  11.2$\pm$3.5   & $6.3\pm0.5$ & $0.295\pm0.092\pm0.028$ & \\
$D^{(*)}$ Modes		&  42.3$\pm$8.4   & $0.9\pm0.1$ & $0.161\pm0.032\pm0.015$ & \\
\hline\hline
\end{tabular}   
\end{center} 
\end{table*} 

	We proceed to use the $\mbc$ distribution from this analysis in combination
with the one for $B_s^*$ in Ref.~\cite{victor_5s} to obtain an improved measurement
of the $B^*_s$ mass. Since those results
used exactly the same data set as in this analysis, the largest systematic error,
the beam energy calibration of ($+4.6\pm2.9$)~MeV~\cite{victor_5s},
cancels out in the (uncorrected) $\mbc$ difference, $\mbc(B^*_s)-\mbc(B^*)$.
The rightmost peak in Fig.~\ref{fig:b2all_mbc} corresponds to $\y5s\to\bstbst$,
and its mean value is measured to be $(5333.1\pm1.3({\rm stat}))$~MeV/$c^2$.
The $\mbc$ peak value for $\y5s\to~B_s^*\bar{B}_s^*$ from Ref.~\cite{bs}
is $(5418.2\pm1.0\pm3.0)$~MeV, where we have added back the ($+4.6\pm2.9$)~MeV 
beam energy correction to obtain an uncorrected value. The difference in
the $\mbc$ peak values, $\mbc(B_s^*\bar{B}_s^*)-\mbc(B^*\bar{B}^*)$, can be
translated into the mass difference, $M(B_s^*)-M(B^*)$
after correcting for the -1.7 (-0.1)~MeV/$c^2$ bias 
that is introduced due to our use of reconstructed $B_{(s)}$ instead of 
$B^*_{(s)}$ mesons. We therefore find a mass
difference $M(B_s^*)-M(B^*)=(86.7\pm1.6\pm0.2)$~MeV/$c^2$. 
The 1.6~MeV/$c^2$ error is statistical and the 0.2~MeV/$c^2$ uncertainty is 
from systematic errors in fitting our $\mbc$ spectrum.
Combining this mass difference with $M(B^*)=(5325.0\pm0.6)$~MeV/$c^2$~\cite{pdg}, we 
obtain an improved value for the $B^*_s$ mass, $M(B^*_s) = (5411.7\pm1.6\pm0.6)$~MeV/$c^2$.
Using the well-measured $B_s$ mass from CDF of 
$M(B_s)=(5366.01\pm0.73\pm0.33)$~MeV/$c^2$~\cite{cdf_bs}, we determine
the $1^{-}-0^{-}$ mass splitting $M(B_s^*)-M(B_s)=(45.7\pm1.7\pm0.7)$~MeV/$c^2$. 
This mass splitting measurement supersedes the previous CLEO result~\cite{victor_5s} of 
$(48\pm1\pm3)$~MeV/$c^2$~\cite{victor_5s} and is significantly
more precise than an earlier indirect measurement of $(47.0\pm2.6)$~MeV/$c^2$~\cite{cusb_bs}.
It is also consistent with the corresponding splitting in the 
$B_d$ system of (45.78$\pm$0.35)~MeV/$c^2$~\cite{pdg} 
as expected from heavy-quark symmetry~\cite{bmass}. 

	In summary, we have measured or set upper limits on the rates for the
various final states in $\y5s$ decay. We find that predictions of the 
UQM~\cite{uqm} are consistent with our findings that 
$\bstbst$ is dominant, with a measured 
value of $(74\pm15)\%$ of the total $B$ rate. The $\bbstar$ rate is measured to be 
about 1/3 of the $\bstbst$ rate. 
Upper limits on $\bbbar$, $\bother$ and $\bbpipi$ have also been presented. Lastly, we 
utilized the $\mbc$ peak positions for $B^*$ and $B^*_s$~\cite{victor_5s} to 
extract $M(B^*_s) = (5411.7\pm1.6\pm0.6)~{\rm MeV/c^2}$,
which is the most precise value of the $B^*_s$ mass to date.

	We gratefully acknowledge the effort of the CESR staff 
in providing us with excellent luminosity and running conditions.
This work was supported by the A.P.~Sloan Foundation,
the National Science Foundation, and the U.S. Department of Energy.


\begin{thebibliography}{99}
\bibitem{hel_angle}
For this decay, the helicity angle $\theta_h$ is the angle between the $\pi^0$ direction
in the  $\rho$ rest frame and the $\rho$ direction in the $D$ rest frame.
\bibitem{mdif}
The $B^*$-$B$ mass difference is $45.78\pm0.35$ MeV/$c^2$, to which a 1.7 MeV/$c^2$ 
Lorentz boost correction was applied that reflects our use of the $B$ meson
momentum in computing $\mbc$ rather than the $B^*$ momentum.
\bibitem{bs}
This value is reduced by the beam energy correction of 4.6~MeV which was applied
in Ref.~\cite{victor_5s}.
\end{thebibliography}
\end{document}